\documentclass[twocolumn,showpacs,amsmath,amssymb,prl]{revtex4}
\usepackage{graphicx}% Include figure files
\usepackage{soul}

\begin{document}
\title{Thermal conductivity of one-dimensional lattices with self-consistent heat baths: a heuristic derivation}

\author{Nianbei Li$^{1,2}$}
\author{Baowen Li$^{1,3}$}
\email{phylibw@nus.edu.sg}

\affiliation{$^1$ Department of Physics and Centre for
Computational Science and Engineering, National University of
Singapore, Singapore 117542,
 Republic of Singapore\\
$^2$ Institut f\"ur Physik, Universit\"at Augsburg,
Universit\"atsstr. 1, D-86135 Augsburg, Germany\\
$^3$ NUS Graduate School for Integrative Sciences and Engineering,
Singapore 117597, Republic of Singapore}

\date{4 Feb 2009}

\begin{abstract}
We derive the thermal conductivities of one-dimensional harmonic and
anharmonic lattices with self-consistent heat baths (BRV lattice)
from the Single-Mode Relaxation Time (SMRT) approximation. For
harmonic lattice, we obtain the same result as previous works.
However, our approach is heuristic and reveals phonon picture {\it
explicitly} within the heat transport process. The results for
harmonic and anharmonic lattices are compared with numerical
calculations from Green-Kubo formula. The consistency between
derivation and simulation strongly supports that effective
(renormalized) phonons are energy carriers in anharmonic lattices
although there exist some other excitations such as solitons and
breathers.

\end{abstract}
 \maketitle

Heat conduction exhibits diversified behaviors for one-dimensional
lattices in terms of heat current $j$ as the function of lattice
length $N$\cite{Review1,Review2}. Harmonic lattice possesses
ballistic heat transport\cite{HL} in which the heat current is
independent of the lattice length, $j \sim N^0$. Anharmonic lattices
without external potential such as FPU-like lattices have anomalous
heat conduction\cite{FPU} in the sense $j \sim N^{-\alpha}$, where
$0<\alpha<1$. Anharmonic lattices with external potential such as
Frenkel-Kontorova (FK) and $\phi^4$ lattices show normal Fourier's
heat conduction\cite{FK,phi4} since $j \sim N^{-1}$. After intensive
numerical simulations over the last decade, it has been commonly
accepted that anharmonicity and external potential are sufficient to
induce Fourier's heat conduction law. In fact, Bolsterli, Rich and
Visscher (BRV) have obtained in 1970 the Fourier's heat conduction
by attaching external heat baths to every atom of a harmonic
chain\cite{BRV}. In this so-called BRV lattice, the interaction
between heat baths and harmonic chain resembles the external
anharmonic forces.

The BRV lattice plays a fundamental role in understanding the
Fourier's law in one-dimensional lattices as exemplified by recent
studies\cite{Bonetto, Dhar}. In Ref.\cite{Bonetto}, Bonetto,
Lebowitz and Lukkarinen revisit the thermal transport of BRV lattice
with a mathematically rigorous treatment. In addition, they also
obtain the same result from the derivation of Green-Kubo formula. In
Ref.\cite{Dhar}, they deal with the quantum BRV lattice and obtain
the exact result in the classical (high-temperature) limit. However,
the phonon picture during thermal transport is not reflected {\it
explicitly} in their approaches. Harmonic lattice only allows
phonons as the collective motion. There is no doubt that phonons
should be the energy carriers for harmonic lattice. However for
anharmonic lattices, there are some other collective motions such as
solitons\cite{Solitons} and breathers\cite{Breathers} existing as
the consequence of anharmonicity. They are proposed as the energy
carriers responsible for some novel transport phenomena, e.g.
thermal rectifying in asymmetric nanotubes\cite{Chang}, anomalous
heat conduction in FPU-$\beta$ lattice\cite{Aoki} and so
on\cite{Solitonheat,Breatherheat}. An investigation for anharmonic
BRV lattices where external heat baths are attached to every atom of
an anharmonic chain is desirable and this may shed some light on the
role of energy carriers for anharmonic lattice.

In this work, we study the thermal conductivities of  BRV lattices
from a different approach named as the Single-Mode Relaxation Time
(SMRT) approximation from Boltzmann equation\cite{SMRT1,SMRT2}. Both
harmonic and anharmonic BRV lattices have been considered. The
results from SMRT approximation are also compared with numerical
simulations from Green-Kubo method. The harmonic BRV lattice is
described as a one-dimensional harmonic chain of $N$ atoms attached
by independent Langevin heat baths. The Hamiltonian of this harmonic
chain reads
\begin{equation}\label{ham}
H=\sum^{N}_{i=1}\left[\frac{1}{2}p^{2}_{i}+\frac{1}{2}
\omega^{2}(q_{i+1}-q_{i})^2+\frac{1}{2}\gamma^2q^2_i\right]
\end{equation}
where $q_i$ is the $i$'th atom's displacement from its equilibrium
position. $\omega^2$ and $\gamma^2$ represent the strength of
inter-atom and external potential respectively. For a chain with
lattice constant $a$ and atom mass $m$, we adopt the dimensionless
units by setting $a=m=1$. A periodic boundary condition
$q_{N+1}=q_{1}$ is used. The heat bath attached to every atom is
modeled by stochastic Langevin heat bath. The coupling strength is
determined by the friction coefficient $\lambda$ in Langevin
dynamics. Thus the equation of motion of the harmonic BRV lattice is
expressed as
\begin{equation}\label{i-space}
\ddot{q}_i=-\frac{\partial{H}}{\partial{q_i}}+\xi_i-\lambda\dot{q}_i
\end{equation}
where $\xi_i$ is the Gaussian white noise with $\langle
\xi_i(t)\rangle=0$ and $\langle \xi_i(t)\xi_i(0)\rangle=2\lambda
k_{B}T\delta(t)$.

According to the SMRT approximation\cite{SMRT1,SMRT2}, the thermal
conductivity for one-dimensional system has a compact expression
\begin{equation}\label{kappa}
\kappa=\frac{c}{\pi}\int^{\pi}_{0}v^2(k)\tau(k)dk
\end{equation}
where $c$ is the specific heat, $k$ the phonon wavevector, $v(k)$
the group velocity, and $\tau(k)$ the phonon relaxation time of mode
$k$. We focus on the classical transport processes where specific
heat $c=k_{B}$ for harmonic lattice with Hamiltonian Eq.
(\ref{ham}).

To obtain the group velocity $v(k)$, we need to find the phonon
dispersion relation. For Eq. (\ref{ham}), one readily obtains the
following dispersion relation
$\omega_k=\omega\sqrt{4\sin^2{k/2}+\nu^2}$ for $k\in[0,\pi]$ with
$\nu^2\equiv \gamma^2/\omega^2$. Thus the phonon group velocity is
obtained
\begin{equation}\label{gv}
v(k)=\frac{\partial{\omega_k}}{\partial{k}}
=\omega\frac{\sin{k}}{\sqrt{4\sin^2{\frac{k}{2}}+\nu^2}}
\end{equation}
If the system has no external potential, i.e. $\gamma=0$, the phonon
group velocity is $v(k)=\omega\cos{k/2}$. The existence of external
potential shifts the phonon band off the zero point and reduces the
phonon group velocity, as can be seen from Eq. (\ref{gv}).

To analyze the phonon relaxation time $\tau(k)$, we consider the
velocity auto-correlation function $\langle
\dot{q}_k(t)\dot{q}_k(0)\rangle$ for mode $k$. Here $q_k$ is the
counterpart of $q_i$ in mode space after some canonical
transformation $q_k=\sum^{N}_{i=1}S_{ki}q_i$. $S$ is the
transformation matrix. We use symbol $k$ to represent wavevector and
mode index simultaneously since it will not cause confusion. After
canonical transformation, the Hamiltonian can be decomposed into
separated harmonic oscillators specified by a frequency $\omega_k$.
Performing the transformation on both sides of Eq. (\ref{i-space}),
we obtain the separated equation of motion in mode space:
$\ddot{q}_k=-\omega^2_{k}q_k+\xi_k-\lambda\dot{q}_k$. After
transformation, $\xi_k$ still obeys the statistical properties of
Gaussian white noise. Especially, the statistical properties of the
fluctuations will not depend on the choice of the potential form.
The above Langevin equation can be solved\cite{Hanggi} by setting
the first term on the right hand side to zero:
$\dot{q}_k(t)=\dot{q}_k(0)e^{-\lambda
t}+\int^{t}_{0}dt'e^{-\lambda(t-t')}\xi_k(t')$. Substituting the
solution of $\dot{q}_k(t)$ into the velocity auto-correlation
function, we have
\begin{equation}
\langle\dot{q}_k(t)\dot{q}_k(0)\rangle=\langle\dot{q}^2_k(0)\rangle
e^{-\lambda t}
\end{equation}
The ensemble average of the terms containing $\xi_k(t)$ vanish due
to the statistical property of Gaussian white noise. The velocity
auto-correlation function decays exponentially with a characteristic
time of $1/\lambda$. Therefore we come to the conclusion that the
phonons acquire a frequency-independent relaxation time
$\tau(k)=1/\lambda$ due to the coupling with the external Langevin
heat bath.

As a result, the thermal conductivity can be obtained from Eq.
(\ref{kappa})
\begin{eqnarray}\label{kappa-har}
\kappa&=&\frac{k_{B}}{\pi}\int^{\pi}_{0}\left(\frac{\omega\sin{k}}{\sqrt{4\sin^2{\frac{k}{2}}+\nu^2}}\right)^2\frac{1}{\lambda}dk
\nonumber\\
&=&\frac{k_{B}\omega^2}{\lambda[2+\nu^2+\sqrt{\nu^2(4+\nu^2)}]}
\end{eqnarray}
Thus we have obtained the same exact result as in previous
works\cite{BRV,Bonetto,Dhar} from a heuristic derivation. The phonon
picture for the thermal transport is {\it explicitly} revealed in
our derivation.

\begin{figure}
\includegraphics[width=\columnwidth]{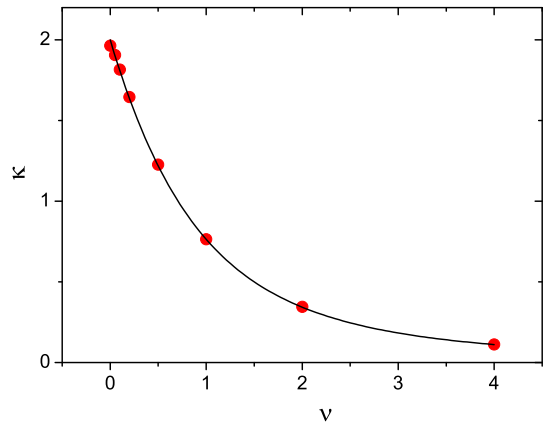}
\vspace{-.5cm} \caption{(Color-online) Thermal conductivity $\kappa$
vs $\nu$. Solid curve is from Eq. (\ref{kappa-har}), and bullets are
numerical results from Green-Kubo method. The parameters are
$\omega=1, \lambda=0.25, T=1$ for a lattice with $N=100$. The
Boltzmann constant $k_{B}$ is set to unity.} \label{fig:kappa-nu}
\end{figure}

As we have mentioned in the introduction, it has been proved that
the Green-Kubo formula renders the same result\cite{Bonetto}. Here
we calculate the thermal conductivities of harmonic BRV lattice by
Green-Kubo formula numerically. The advantage of numerical
Green-Kubo method lies in the fact that it can be easily extended to
anharmonic BRV lattice where no existing theoretical analysis has
ever been applied. The thermal conductivity is calculated
from\cite{Bonetto}
\begin{equation}
\kappa=\frac{1}{k_{B}T^{2}N}\int^{\infty}_{0}\langle J(t)J(0)\rangle
dt
\end{equation}
where $J(t)=\sum^{N}_{i=1}j_i$ and
$j_i\equiv\dot{q}_{i}\frac{\partial{V(q_i,q_{i+1})}}{\partial{q_i}}$
is the heat current within the chain. $V(q_i,q_{i+1})$ represents
the inter-atom potential. The length and temperature independence of
$\kappa$ has been verified by Green-Kubo simulations (not shown
here). In Fig.\ref{fig:kappa-nu}, we plot the thermal conductivity
as a function of $\nu$ at $\omega=1,\lambda=0.25,T=1$ for a length
$N=100$. The numerical result for a 100-atoms chain has already
approached the continuous-limit result. There is no surprise since
the mean-free-path $l(k)$ of phonons ($l(k)=v(k)\tau(k)\sim
\omega/\lambda \sim 4$) is much shorter than the chain length
$N=100$.

Next we shall consider the anharmonic BRV lattice. As we have seen
above, the existence of external potential only affects the group
velocity. For simplicity, we only consider the anharmonic BRV
lattice without external potential term. First we consider the
anharmonic case containing FPU-$\beta$ chain. The Hamiltonian of
FPU-$\beta$ chain is
\begin{equation}\label{fpub}
H=\sum^{N}_{i=1}\left[\frac{1}{2}p^{2}_{i}+\frac{1}{2}(q_{i+1}-q_{i})^2
+\frac{1}{4}(q_{i+1}-q_{i})^4\right]
\end{equation}
Due to the anharmonic interaction, the canonical transformation of
Eq. (\ref{i-space}) yields coupled equations of motion in mode
space. In this case, we need to evoke the effective phonon
theory\cite{Lnb}, which considers the phonon-phonon interaction as a
kind of mean field. The anharmonic Hamiltonian can be viewed as
weakly coupled effective phonons with a renormalized frequency
$\hat{\omega}_k$. Thus by ignoring the weakly coupled interactions
within effective phonons, we obtain the approximate equations of
motion in mode space:
$\ddot{q}_k=-\hat{\omega}^2_{k}q_k+\xi_k-\lambda\dot{q}_k$. The
statistical properties of the fluctuations don't depend on the
choice of the potential form. With similar analysis, the relaxation
time for effective phonons is still the frequency-independent
$1/\lambda$. The major anharmonic effect comes from the group
velocity of effective phonons. According to the effective phonon
theory, the effective phonon frequency $\hat{\omega}_k$ in
FPU-$\beta$ chain is proportional to the phonon frequency
$\omega_k=2\sin{k/2}$ with a temperature(anharmonicity)-dependent
prefactor, i.e. $\hat{\omega}_k=\sqrt{\alpha}\omega_k$. The
temperature-dependent coefficient $\alpha$ has the following
analytic expression
$\alpha=1+\int^{\infty}_{0}\phi^{4}e^{-\beta(\phi^2/2+\phi^4/4)}d\phi/\int^{\infty}_{0}\phi^{2}e^{-\beta(\phi^2/2+\phi^4/4)}d\phi$
with $\beta\equiv 1/k_{B}T$. From this, we obtain the group velocity
of effective phonons:
$v(k)=\partial{\hat{\omega}_k}/\partial{k}=\sqrt{\alpha}\cos{k/2}$.

\begin{figure}
\includegraphics[width=\columnwidth]{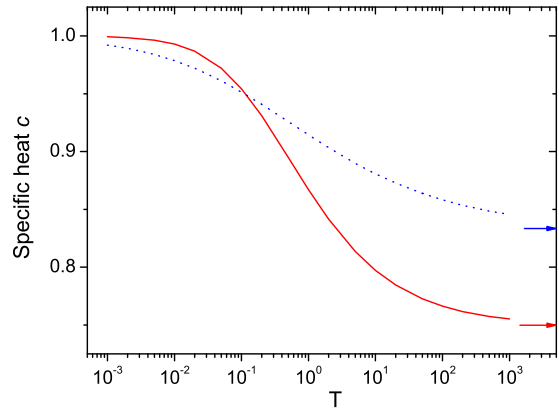}
\vspace{-1cm} \caption{(Color-online) Specific heat $c$ of the FPU
lattices as a function of temperature $T$. Solid curve is for
FPU-$\beta$ chain and dotted curve is for symmetric FPU-$\alpha$
chain. The high temperature limit values of specific heat, $3/4$ for
FPU-$\beta$ and $5/6$ for symmetric FPU-$\alpha$ chain (see Eq.
(\ref{fpua})), are marked by two arrows pointing to right. Boltzmann
constant $k_{B}$ has been set to unity.} \label{fig:specificheat}
\end{figure}

The anharmonic interaction will also make the specific heat a
function of temperature. This effect cannot be neglected when
temperature is not very low. For a chain with $N$ atoms, the
specific heat is determined by $c\equiv\langle H\rangle /NT$. As a
result of energy equipartition theory, the contribution to specific
heat from kinetic energy is half of the Boltzmann constant
$k_{B}/2$. The contribution from potential energy can be derived by
using the equality $Nk_{B}T=\sum^{N}_{i=1}\langle
q_{i}\frac{\partial{H}}{\partial{q_i}}\rangle$. Substituting the
Hamiltonian, Eq. (\ref{fpub}), into the above equality and using the
definition of specific heat, one can obtain the specific heat for
FPU-$\beta$
\begin{equation}\label{c-fpub}
c=k_{B}\left[1-\frac{\int^{\infty}_{0}\frac{\phi^{4}}{4}e^{-\beta(\frac{\phi^2}{2}+\frac{\phi^4}{4})}d\phi}{T\int^{\infty}_{0}e^{-\beta(\frac{\phi^2}{2}+\frac{\phi^4}{4})}d\phi}\right]
\end{equation}
The low and high temperature limit value of $c$ are $k_{B}$ and
$\frac{3}{4}k_{B}$, respectively. The temperature-dependent $c$ is
plotted in Fig.\ref{fig:specificheat}. The monotonically decreasing
specific heat as the function of temperature looks counter-intuitive
as we all know that the specific heat of real material is a
monotonically increasing function of temperature. We have to
emphasize two things clearly here. First we only deal with classical
specific heat. For harmonic chain, the classical specific heat is a
temperature-independent constant $k_B$. Second, we have introduced
the anharmonic interaction. The anharmonic interaction actually
reduces the specific heat. Thus the "counter-intuitively" decreasing
specific heat with temperature is a classically
anharmonicity-induced result.

\begin{figure}
\includegraphics[width=\columnwidth]{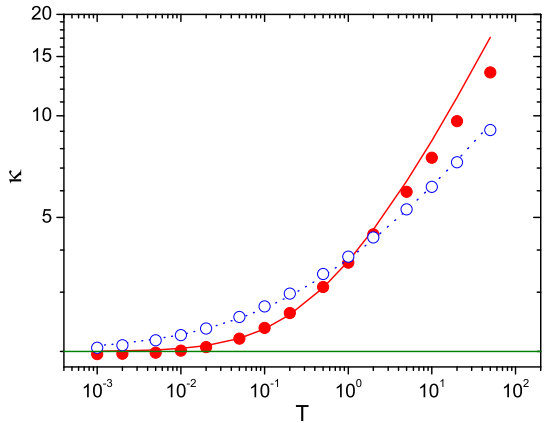}
\vspace{-1cm} \caption{(Color-online) Temperature-dependent  thermal
conductivity $\kappa$ of anharmonic BRV lattices. Solid and dotted
curves are from Eq. (\ref{kappa-fpub}) for FPU-$\beta$ and symmetric
FPU-$\alpha$ BRV lattices, respectively. Solid circles and hollow
circles are the numerical Green-Kubo results for FPU-$\beta$ and
symmetric FPU-$\alpha$ BRV lattices, respectively. The horizontal
line marks the value of $\kappa$ for harmonic BRV lattice. The
simulations are performed at $\lambda=0.25$ for a lattice with
$N=100$. Boltzmann constant $k_{B}$ has been set to unity.}
\label{fig:kappa-T}
\end{figure}

The thermal conductivity of FPU-$\beta$ BRV lattice can be derived
from Eq. (\ref{kappa}) as
\begin{equation}\label{kappa-fpub}
\kappa=\frac{c}{\pi}\int^{\pi}_{0}
\left(\sqrt{\alpha}\cos{\frac{k}{2}}\right)^2\frac{1}{\lambda}dk
=\frac{c\alpha}{2\lambda}
\end{equation}
Thus, $\kappa$ depends on temperature via the product of $c$ and
$\alpha$. To verify this prediction, we perform the Green-Kubo
calculations of $\kappa$. The results are plotted in
Fig.\ref{fig:kappa-T}. Solid curve is our prediction of Eq.
(\ref{kappa-fpub}) and red solid circles are the numerical values of
$\kappa$. Our prediction agrees very well with the Green-Kubo
calculations for a very wide temperature range of about four orders
of magnitudes.

It must be emphasized that our prediction is derived only by
assuming effective phonons as the energy carriers and ignoring the
weak interactions between them. The consistency between theoretical
analysis and numerical simulations clearly demonstrate that it is
effective phonons, a kind of renormalized phonons due to anharmonic
interactions, carry heat energy in thermal transport processes for
FPU-$\beta$ lattice.

Besides FPU-$\beta$ BRV lattice, we also consider another simple
anharmonic BRV lattice containing a symmetric FPU-$\alpha$ chain
\begin{equation}\label{fpua}
H=\sum^{N}_{i=1}\left[\frac{1}{2}p^{2}_{i}+\frac{1}{2}(q_{i+1}-q_{i})^2
+\frac{1}{3}\left|q_{i+1}-q_{i}\right|^3\right]
\end{equation}
The expression of specific heat $c$ and coefficient $\alpha$ are
obtained
\begin{eqnarray}\label{fpua-c}
c&=&k_{B}\left[1-\frac{\int^{\infty}_{0}\frac{\phi^{3}}{6}e^{-\beta(\frac{\phi^2}{2}+\frac{\phi^3}{3})}d\phi}
{T\int^{\infty}_{0}e^{-\beta(\frac{\phi^2}{2}+\frac{\phi^3}{3})}d\phi}\right]\nonumber\\
\alpha&=&1+\frac{\int^{\infty}_{0}\phi^{3}e^{-\beta(\frac{\phi^2}{2}+\frac{\phi^3}{3})}d\phi}{\int^{\infty}_{0}\phi^{2}e^{-\beta(\frac{\phi^2}{2}+\frac{\phi^3}{3})}d\phi}
\end{eqnarray}
As we have discussed for FPU-$\beta$ BRV lattice, the symmetric
FPU-$\alpha$ BRV lattice also has a frequency-independent phonon
relaxation time, i.e. $\tau(k)=1/\lambda$. The final expression of
$\kappa$ is also the same as Eq. (\ref{kappa-fpub}). Here we should
keep in mind that the specific heat $c$ and $\alpha$ are now coming
from Eq. (\ref{fpua-c}). This prediction is plotted in Fig.
\ref{fig:kappa-T} as the dotted curve. The numerical results (hollow
circles in Fig. \ref{fig:kappa-T}) are in good agreement with this
prediction.

In summary, we have derived analytically the thermal conductivities
of harmonic and anharmonic BRV lattices from the SMRT approximation.
The derivation is heuristic and the phonon (effective phonon)
picture is {\it explicit} during derivation. For harmonic BRV
lattice, we obtain the same exact result as in previous works. For
anharmonic BRV lattices, we obtain the approximate results and
compare them with numerical simulations from the Green-Kubo formula.
The consistency between our theoretical results and numerical
simulations demonstrates that the effective (renormalized) phonons
should be the fundamental energy carriers of anharmonic lattices.
The contributions from solitons and breathers, if any, are
negligible at least for FPU-like lattices.

We thank Pawl Keblinski for useful suggestions. This work is
supported by grant R-144-000-203-112 from the Ministry of Education
of the Republic of Singapore.

\end{document}